\documentclass[twocolumn,aps,amssymb]{revtex4}
\usepackage{graphicx}

\begin{document}
\title{Negative Refraction in Ferromagnet/Superconductor Superlattices}
\author{A. Pimenov}
\affiliation{Experimentalphysik V, Center for Electronic
Correlations and Magnetism, Universit\"{a}t Augsburg, 86135
Augsburg, Germany}
\author{P. P. Przyslupski}
\affiliation{Institute of Physics, Polish Academy of Sciences,
02-668 Warszawa, Poland}
\author{B. Dabrowski}
\affiliation{Department of Physics, Northern Illinois University,
60115 DeKalb, Illinois, USA}
\author{A. Loidl}
\affiliation{Experimentalphysik V, Center for Electronic
Correlations and Magnetism, Universit\"{a}t Augsburg, 86135
Augsburg, Germany}

\date{\today}

\begin{abstract}
Negative refraction, which reverses many fundamental aspects of
classical optics, can be obtained in systems with negative
magnetic permeability and negative dielectric permittivity. This
Letter documents an experimental realization of negative
refraction at millimeter waves, finite magnetic fields and
cryogenic temperatures utilizing a multilayer stack of
ferromagnetic and superconducting thin films. In the present case
the superconducting YBa$_2$Cu$_3$O$_7$ layers provide negative
permittivity while negative permeability is achieved via
ferromagnetic (La:Sr)MnO$_3$ layers for frequencies and magnetic
fields close to the ferromagnetic resonance. In these
superlattices the refractive index can be switched between
positive and negative regions using external magnetic field as
tuning parameter.
\end{abstract}


\maketitle

Over the past centuries lenses were important tools to discover
the fundamental laws of optics and even to develop an early
understanding of the universe. Since then diffraction and
reflection are well understood and belong to the basics of
fundamental physics. Classical optics also sets a lower limit for
the resolution of any optical device: The focus is limited by the
wave length of light. However, in 1968 Veselago \cite{veselago}
proposed that a material with a negative index of refraction
(NIR), i. e. with electric permittivity $\varepsilon  < 0$ and
magnetic permeability $\mu < 0$, would reverse all known optical
properties \cite{pendry}. Later on, the conditions leading to NIR
have been extended to media with dissipation \cite{depine}. In a
negative refraction material (NRM) electromagnetic waves propagate
in a direction opposite to that of the flow of energy. 30 years
later this predictions have experimentally been confirmed
\cite{smith,shelby,par,houck,fot,cub,parimi}. Unfortunately,
nature does not provide an adequate material and hence,
metamaterials  \cite{smith,shelby,par,houck} and photonic crystals
\cite{fot,cub,parimi} have been utilized as NRMs. Metamaterials
are well-designed arrays of metallic posts yielding $\varepsilon <
0$, interdispersed with an array of split-ring resonators
revealing an effective negative magnetic permeability ($\mu < 0$).
Photonic crystals are two- or three-dimensional arrays of periodic
arrangements of dielectric or metallic elements.

One of the most outstanding predictions for materials with a NIR
has been made by Pendry \cite{pendry}: a NRM reveals a resolution
beyond the diffraction limit and allows the construction of ideal
lenses. Also this superlensing effect has been experimentally
verified for a number of metamaterials at microwave frequencies
\cite{lagarkov,grbic} and very recently even for optical
frequencies \cite{fang}.

In this Letter we demonstrate experimentally that a superlattice
designed of subsequent thin ferromagnetic and superconducting
layers also provides a negative index of refraction. In this
metamaterial the superconducting layers of high-$T_c$
YBa$_2$Cu$_3$O$_7$ (YBCO) provide the negative permittivity, while
the ferromagnetic layers of the insulating manganite
La$_{0.89}$Sr$_{0.11}$MnO$_3$ (LSMO) reveal negative magnetic
permeability close to the ferromagnetic resonance. This
investigated structure is similar to the bilayer system suggested
by Al\`{u} and Engheta \cite{alu}, which has been analyzed in
details by Lakhtakia and Krowne \cite{krowne} showing the
possibility of negative refraction.

LSMO/YBCO superlattices \cite{piotr2} were deposited on (100)
(LaAlO$_3$)$_{0.3}$(Sr$_2$LaAlTaO$_6$)$_{0.7}$ (LSAT) substrates
by multitarget high-pressure sputtering \cite{piotr1}. Two targets
with nominal composition of YBa$_2$Cu$_3$O$_7$ and
La$_{0.98}$Sr$_{0.11}$MnO$_3$ were used for deposition. The
thickness of different layers was controlled by the deposition
times of the respective targets. Two samples from these targets
have been prepared, with compositions [66 unit cells (uc) LSMO / 8
uc YBCO]$_8$ and [33 uc LSMO / 8 uc YBCO]$_8$. The observed
properties of both compositions were qualitatively similar, except
for a weaker magnetic signal of the latter sample (roughly a
factor of two due to the lower LSMO content). Therefore, in the
following only the results from the first sample will be shown.
The superlattices were characterized using X-ray and
SQUID-magnetization measurements. For both samples a clear
diamagnetic signal (Meissner effect) has been observed at low
temperatures. Characteristic susceptibility curves in the
field-cooled (FC) and zero-field-cooled regime (ZFC) are shown in
the upper panel of Fig. 1. From these data the ferromagnetic
Curie-temperature $T_{\rm C} \simeq 207$ K has been derived and
from the magnetization curves the effective magnetic moment per
manganese ion can be estimated as $M_{LSMO} = 1.5 \mu_B$, $\mu_B$
denoting the Bohr's magneton.

The dynamic experiments in the frequency range 0.06 THz $< \nu <$
0.5 THz have been carried out in transmittance experiments using a
Mach-Zehnder interferometer \cite{pronin}. Temperature- and
magnetic field-dependent experiments have been carried out in a
split-coil magnet providing magnetic fields up to 7 T and
temperatures 1.8 K $\leq T \leq$ 300 K. This arrangement allows to
investigate the transmittance and phase shift of the plane
parallel samples as function of frequency, temperature, and
external magnetic field. The effective permittivity and
permeability of the superlattice
have been calculated from these quantities using the Fresnel
optical equation for the complex transmission coefficient of the
plane-parallel plate \cite{born}:
\begin{equation}\label{eqtran}
    t=\frac{(1-r^{2})t_{1}}{1-r^{2}t_{1}^{2}}
\end{equation}
Here $r=(\sqrt{\varepsilon^* /\mu^* }-1)/(\sqrt{\varepsilon^*
/\mu^* }+1)$ is is the reflection amplitude at the air-sample
interface, $t_1=\exp (-2\pi i\sqrt{\varepsilon^* \mu^*
}\,d/\lambda)$ is the ``pure'' transmission amplitude,
$\varepsilon^*$ and $\mu^*$ are the (complex) dielectric
permittivity and magnetic permeability
 of the sample, respectively, $d$ is the sample
thickness, and $\lambda$ the radiation wavelength. The effective
conductivity of the superlattices is obtained via
$\sigma^*=\sigma_1+i\sigma_2=\varepsilon_0\varepsilon^*\omega/i$,
where $\varepsilon_0$ is the permittivity of vacuum and  $\omega =
2\pi\nu$ the angular frequency. Eq. (\ref{eqtran}) is written in
the approximation neglecting the properties of the substrate.
During the experiments the full expression \cite{born} has been
used, which in most cases leads to results closely similar to Eq.
(\ref{eqtran}) and is omitted here for simplicity. The properties
of the LSAT substrate were measured in a separate experiment and
could well be approximated by a frequency-independent refractive
index $n \simeq 4.8$.

In the THz frequency range and in the absence of the external
magnetic field both YBCO and LSMO are non-magnetic in good
approximation and therefore $\mu^*=1$ can be used in calculations
for $H=0$. Substantial deviations from $\mu^*=1$ can be expected
in the vicinity of the ferromagnetic mode in LSMO only \cite{spin}
and are strongly magnetic field dependent. On the contrary, the
dielectric permittivity is only weakly field-dependent and can be
accounted for using an additional linear term. This procedure
allows to separate the contributions of the dielectric
permittivity and magnetic permeability in Eq. (\ref{eqtran})
combining zero-field data with the results in external magnetic
fields.

\begin{figure}[]
\centering
\includegraphics[width=8cm,clip]{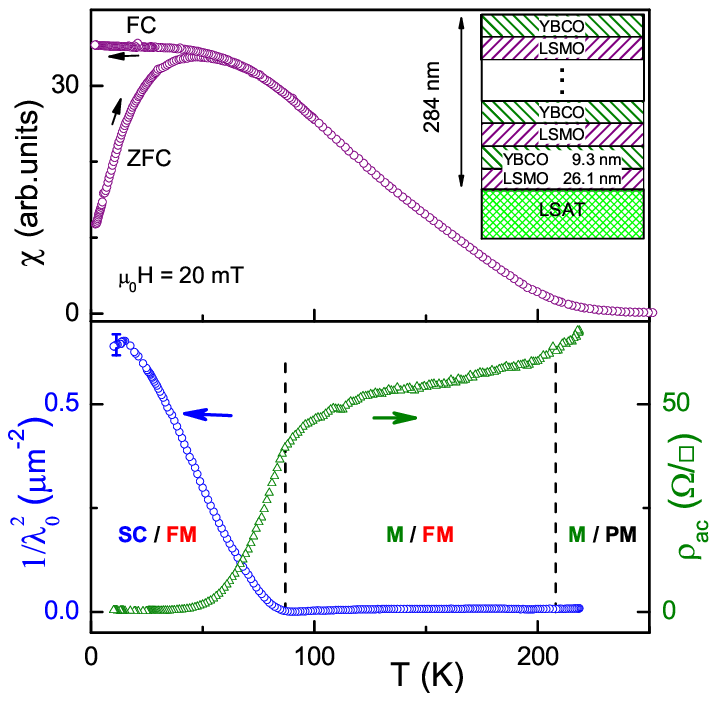}
\caption{ Upper panel: Magnetic susceptibility of the LSMO/YBCO
superlattice measured in the field-cooling (FC) and
zero-field-cooling regime (ZFC). The inset shows schematically the
composition of the sample. \\ Lower panel: Right scale:
resistivity of the superlattices at a frequency of 90 GHz. Left
scale: superconducting penetration depth. Dashed lines indicate
the transition temperatures between different phases, SC -
superconducting, M - metallic, FM - ferromagnetic, and PM -
paramagnetic.} \label{frho}
\end{figure}

Lower panel of Fig. \ref{frho} shows the effective dynamic
resistivity $\rho_{ac} = $ Re$(1/\sigma^*d)$ of the LSMO/YBCO
superlattice measured at a frequency $\nu=$ 90 GHz. The
superlattice has a total thickness of d = 284 nm. In the
normally-conducting state, $T
>$ 90 K, the resistance of the superlattice is only weakly
temperature dependent and is determined by the metallic
conductivity of YBCO. The real part of the dynamic conductivity
$\sigma_1$ reveals no measurable frequency dependence in the
frequency range of the experiments and the imaginary part is zero
within experimental uncertainties. Both properties are
characteristic for genuine metallic behavior and have been
observed in pure YBCO films using the same technique
\cite{pimenov}. The microwave resistivity is rapidly suppressed
below the superconducting transition temperature $T_c \simeq$ 87
K, qualitatively resembling the behavior of the dc resistance (not
shown). The left scale of the lower panel of Fig. \ref{frho} shows
the superconducting penetration depth of the superlattice,
obtained via $\lambda^{-2} = \mu_0 \omega \sigma_2$. Here $\mu_0$
is the permeability of vacuum. At low temperatures the imaginary
part of the conductivity is inversely proportional to the
frequency, which warrants the frequency-independence of the
penetration depth and is the characteristic feature of the
superconducting state.

\begin{figure}[]
\centering
\includegraphics[width=7cm,clip]{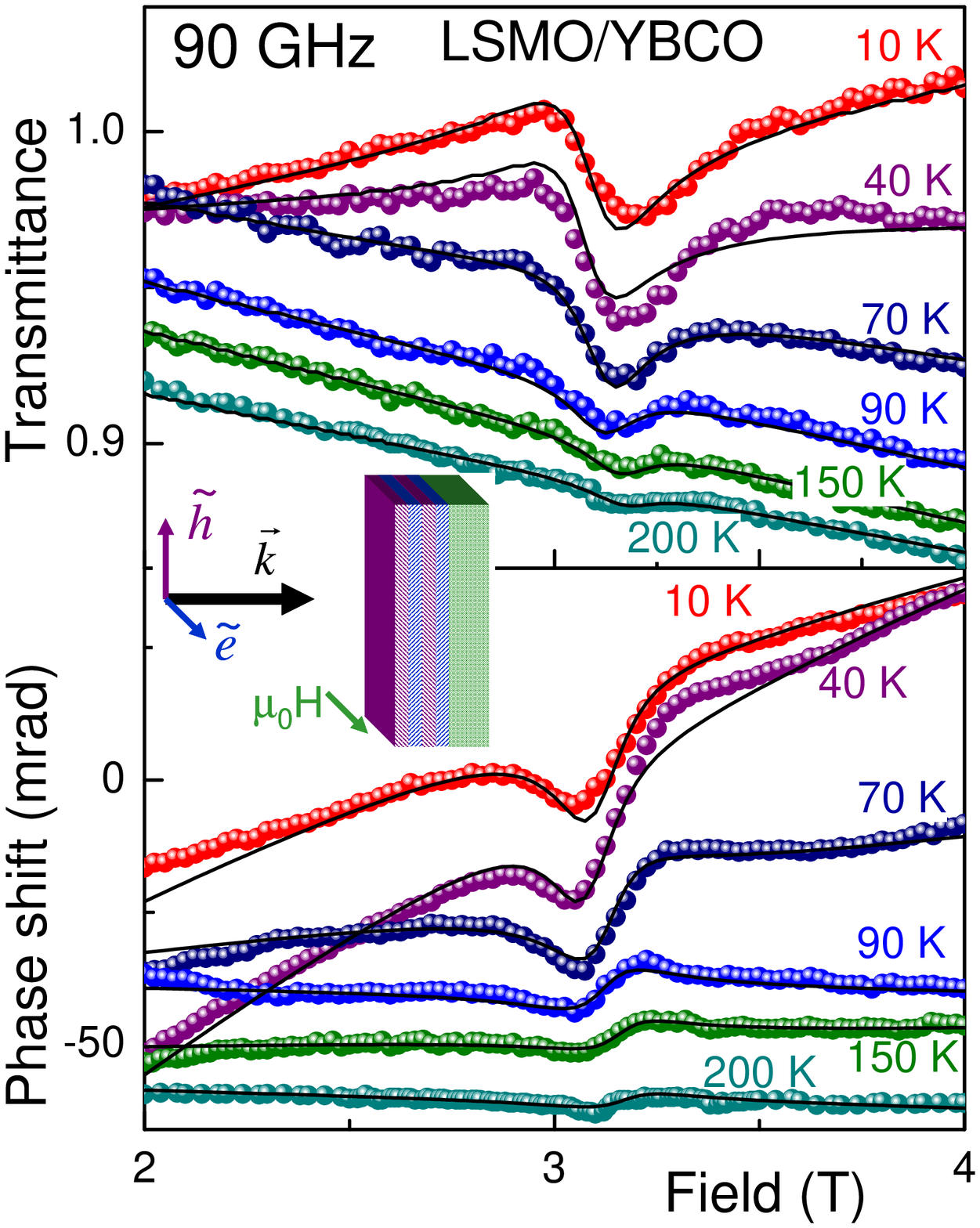}
\caption{Magnetic field-dependence of the transmittance (upper
panel) and the phase-shift (lower panel) of the LSMO/YBCO
superlattices at $\nu =$ 90 GHz and for different temperatures.
Symbols - experiment, lines - fits to the Lorentz-like form of the
ferromagnetic resonance close to 3.1 T. The inset shows the
geometry of the experiment.} \label{ftran}
\end{figure}

Figure \ref{ftran} represents the magnetic field-dependence of the
relative transmittance (upper panel) and the phase shift (lower
panel) of the LSMO/YBCO superlattice at a frequency of 90 GHz. A
ferromagnetic resonance (FMR) close to $\mu_0 H =$ 3.1 T is
clearly observed for all temperatures. Only at $T =$ 200 K the
intensity of the observed mode is just above the noise-level of
the spectrometer. The spectra have been measured in Voigt geometry
\cite{zvezdin}, i.e. with the static magnetic field within the
sample-plane. The geometry of the magnetic field-dependent
experiment is shown as inset in Fig. \ref{ftran}. In addition to
the sharp line of the ferromagnetic resonance, both transmittance
and phase-shift reveal a weak overall field-dependence. This
behavior corresponds to the high-frequency magnetoresistance of
the superconductor/ferromagnet superlattices and is most
pronounced in the superconducting state. The absolute value of
this effect is of the order $|\Delta\sigma|/|\sigma^*| \sim
3\%/{\rm Tesla}$, much weaker than the dc-magnetoresistance with
changes of the order of $1000\%$ \cite{pena}. The high-frequency
magnetoresistance can be formally ascribed to the partial
weakening of superconductivity in the external magnetic field and
is beyond the scope of this work.

The solid lines in Fig. \ref{ftran} have been calculated assuming
a Lorentz shape of the ferromagnetic line, i.e. the magnetic
permeability of a ferromagnet was taken as:
\begin{equation}\label{eqmu}
    \mu^*(H) = \mu_1 + i\mu_2 = 1+ \frac{\Delta\mu H_0 H}{H^2-H_0^2-iH_0\gamma}
\end{equation}
Here $H_0, \gamma$, and $\Delta \mu$ are the resonance field,
width and magnetic contribution of the ferromagnetic line,
respectively. The field dependence of complex conductivity has
been accounted for assuming a linear magnetoresistance effect and
has been obtained from the data far off the resonance position.
Within the experimental uncertainties the width and the position
of the ferromagnetic line are approximately
temperature-independent with $\gamma = 0.2 \pm  0.02$ T and $H_0 =
3.1 \pm 0.02$ T. In the metallic temperature range $T \geq  $ 90 K
the intensity of the line qualitatively follows the magnetization
from $\Delta\mu$(200 K) = 0.2 to   $\Delta\mu$(90 K) = 0.4. This
is in agreement with the expectation of a ferromagnet $\Delta\mu
\sim M/H_0$, where $M$ is the static magnetization. Below the
superconducting transition the intensity decreases with
temperature, which probably results from the screening due to the
onset of the superconductivity.

The form of the FMR line differs qualitatively in the metallic and
the superconducting states. For example, at $T =$ 90 K a symmetric
dip in the transmittance and a resonance-like step in the
phase-shift are fingerprints of a conventional resonance-line. At
low temperature and in the superconducting state the line is
highly asymmetric, and the form of transmittance and phase-shift
curves have interchanged: at $T =$ 10 K a dip appears in the
phase-shift and an asymmetric step in the transmittance.

The strength of the ferromagnetic mode in the superconducting
state leads to negative values of the magnetic permeability close
to the resonance field \cite{chui}. At the same time, the
dielectric permittivity of a superconductor Re$(\varepsilon^*) =
\varepsilon_1 = -\sigma_2/\varepsilon_0\omega$ is large and
negative (here $\mu_0$ is the permittivity of vacuum). Hence, the
situation occurs, where both $\varepsilon_1$ and  $\mu_1$ are
negative, corresponding to the Veselago's criteria for a negative
refractive index. In this case special care has to be taken in
calculating the refractive index and the transmittance of the
sample because the correct sign of the refractive index in the
expression $n=\pm\sqrt{\varepsilon^*\mu^*}$  is not \textit{a
priori} evident. However, due to a continuous transition from the
regions between negative and positive refractive indices, realized
in our experiment, the question of the sign of the square-root is
solved automatically.

\begin{figure}[]
\centering
\includegraphics[width=8.5cm,clip]{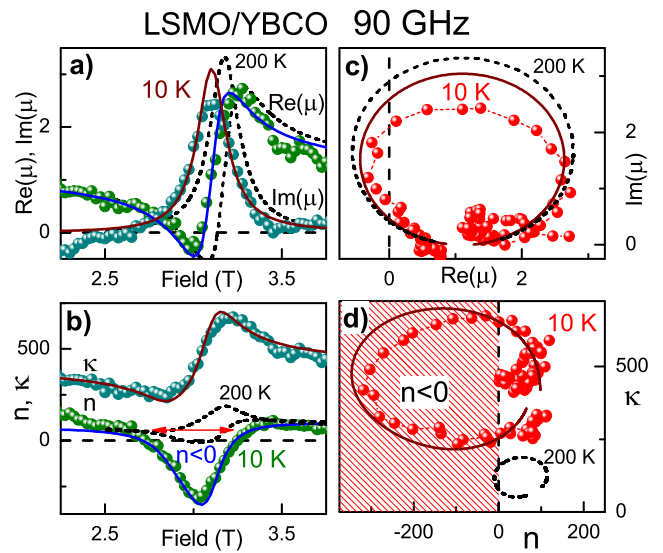}
\caption{Magnetic permeability of LSMO/YBCO superlattices at $T =
10$ K as function of magnetic field (a) and replotted in the
complex-plane (c). The complex refractive index as
field-dependence (b) and in the complex plane (d). The shaded area
corresponds to the region with negative real part of refractive
index. The results for $T =$ 200 K are given for comparison. In
all frames the symbols correspond to experimental data, lines
represent the Lorentz model for the FMR line.} \label{frefr}
\end{figure}

In addition to the fits of the experimental spectra of FMR, the
field dependence of the magnetic permeability and the refractive
index can directly be calculated from the data of Fig.\
\ref{ftran}. The calculated permeability ($\mu_1 =$ Re($\mu^*$),
$\mu_2 =$ Im($\mu^*$)) vs. magnetic field is shown in Fig.\
\ref{frefr}a and the data are re-plotted in the complex plane in
Fig.\ \ref{frefr}c. In this calculation we assumed that the
complex conductivity in the vicinity of the FMR line depends
linearly upon the magnetic field. The symbols in Figs.
\ref{frefr}a-d represent the experimental data which are compared
with the Lorentzian line-shape given by solid lines. As is clearly
documented in Fig.\ \ref{frefr}a, the FMR mode is strong enough to
produce a region with negative permeability between 2.9 T and 3.1
T. In combination with the negative dielectric permittivity of the
superconductor, $\varepsilon^*$(10 K) $= (-13.7 + 6.1i)\cdot
10^4$, this leads to a substantial region with a negative
refractive index (Fig. \ref{frefr}b,d). For comparison, the dashed
lines in Fig. \ref{frefr} show the results obtained at $T =$ 200
K. At this temperature the dielectric permittivity is
characteristic for a metal $\varepsilon^*$(200 K) $= (0.2 +
1.1i)\cdot 10^4$ and the resulting curve of the refractive index
predominantly corresponds to $n > 0$.

Finally, the requirement of cryogenic temperatures and high
magnetic fields hampers possible applications of the
ferromagnet/superconductor multilayers. However, the absolute
values of the external magnetic field can be reduced in the same
system decreasing the frequency of the experiment. Substitution of
a ferromagnet by an antiferromagnet with narrow line of the
antiferromagnetic resonance may completely remove the necessity of
the external magnetic fields. The way to increase the working
temperature of the multilayers could be the substitution of the
superconductor by a good metal revealing negative permittivity.

In conclusion, we have shown experimentally that a region of
negative refractive index can be realized in superlattices of
superconducting and ferromagnetic layers in the superconducting
state close to the ferromagnetic resonance. Compared to
metamaterials based on the metallic rods and split-ring resonators
these results open another way to construct materials with
negative refraction. In the superconductor/ferromagnet
superlattices the refractive index can be switched between
positive and negative regions using an external magnetic field as
a tuning parameter

This work was supported by BMBF (13N6917/0 - EKM), DFG (SFB484 -
Augsburg), US DoE, NSF-DMR-0302617, MaNP, and Polish Committee for
Scientific Research (2006-2008).


\begin{thebibliography}{99}

\bibitem{veselago} V. G.  Veselago, Sov. Phys. Uspekhi \textbf{10}, 509 (1968).

\bibitem{pendry} J. B. Pendry, Phys. Rev. Lett. 85, 3966 (2000).

\bibitem{depine}  R. A. Depine and A. Lakhtakia, Microw. Opt. Technol. Lett. \textbf{41} 315
(2004);  M. W. McCall, A. Lakhtakia, and W. S. Weiglhofer, Eur. J.
Phys. (UK) \textbf{23}, 353 (2002).

\bibitem{smith} D. R. Smith, W. J. Padilla, D. C. Vier, S. C.
Nemat-Nasser, and S.Schultz, Phys. Rev. Lett. \textbf{84}, 4184
(2000).

\bibitem{shelby} R. A. Shelby, D.R. Smith, and S. Schultz, Science \textbf{292}, 77 (2001).

\bibitem{par} C. G. Parazzoli, R. B. Greegor, K. Li, B. E. C.
Koltenbah, and M. Tanielian, Phys. Rev. Lett. \textbf{90}, 107401
(2003).

\bibitem{houck} A. A. Houck, J. B. Brock, and I. L. Chuang, Phys. Rev. Lett. \textbf{90}, 137401 (2003).

\bibitem{fot} S. Foteinopoulou, E. N. Economou, and C. M. Soukoulis, Phys. Rev. Lett. \textbf{90}, 107402 (2003).

\bibitem{cub} E. Cubukcu, K. Aydin, E. Ozbay, S. Foteinopoulou, and C.M. Soukoulis, Nature \textbf{423}, 604 (2003).

\bibitem{parimi} P. V. Parimi, W. T. Lu, P. Vodo, and S. Sridhar, Nature 426, 404 (2003).

\bibitem{lagarkov} A. N. Lagarkov and V. N. Kissel, Phys. Rev. Lett. \textbf{92}, 077401 (2004).

\bibitem{grbic} A. Grbic and G. V. Eleftheriades, Phys. Rev. Lett. \textbf{92}, 117403 (2004).

\bibitem{fang} N. Fang, H. Lee, C. Sun, and X. Zhang, Science \textbf{308}, 534 (2005).

\bibitem{alu} A. Al\`{u} and N. Engheta, IEEE Trans. Antennas
Propag. \textbf{51}, 2558 (2003); J. Appl. Phys. \textbf{97},
093340 (2005).

\bibitem{krowne} A. Lakhtakia and C. M. Krowne, Optik (Germany) \textbf{114}, 305
(2003).

\bibitem{piotr2} P. Przyslupski \textit{et al.}, Phys. Rev. B \textbf{69}, 134428 (2004);
J.Appl.Phys \textbf{95}, 2906(2004).

\bibitem{piotr1} P. Przyslupski, S. Kolesnik, E. Dynowska, S. T. Koskiewicz, and M. Sawicki,
IEEE Trans. Appl. Supercond. \textbf{7}, 2192 (1997).

\bibitem{pronin} A. V. Pronin \textit{et al.}, Phys. Rev. B 57, 14416
(1998); A. Pimenov {\it et al.}, Phys. Rev. B. \textbf{72}, 035131
(2005).

\bibitem{born} M. Born and E. Wolf, \textit{Principles of optics} (Pergamon, Oxford, 1986).

\bibitem{spin} D. Ivannikov {\it et al.}, Phys. Rev. B \textbf{65}, 214422 (2002).

\bibitem{pimenov} A. Pimenov, A. Loidl, G. Jakob, and H. Adrian, Phys. Rev. B \textbf{59}, 4390
(1999); \textbf{61}, 7039 (2000).

\bibitem{zvezdin} A. K. Zvezdin and V. A. Kotov, \textit{Modern Magnetooptics and Magnetooptical Materials}
(Inst. of Physics Publ., Bristol, 1997).

\bibitem{pena} V.Pe\~{n}a \textit{et al.}, Phys. Rev. Lett. 94, 057002 (2005).

\bibitem{chui} S. T. Chui and L. Hu,  Phys. Rev. B \textbf{65}, 144407
(2002).

\end{thebibliography}
\end{document}